\documentclass{iaus}
\usepackage{graphicx}

\title[~~Population Synthesis] %% give here short title %%
{Population Synthesis at the Crossroads}

\author[Claus Leitherer \& Sylvia Ekstr\"om]   %% give here short author list %%
{Claus Leitherer$^1$ \and Sylvia Ekstr\"om$^2$
 }

\affiliation{$^1$Space Telescope Science Institute, 3700 San Martin Dr., Baltimore, MD 21218, USA
 \\ email: {\tt leitherer@stsci.edu}\\[\affilskip]
$^2$Observatoire de Gen\`eve, 51 chemin des Maillettes, CH-1290 Sauverny, Switzerland
 \\ email: {\tt sylvia.ekstrom@unige.ch}
}

\pubyear{2011} %% 
\volume{284}  %% insert here IAU Symposium No.
\pagerange{1--12}
\jname{The Spectral Energy Distribution of Galaxies}
\editors{R.J. Tuffs \&  C.C.Popescu, eds.}
\begin{document}

\maketitle

\begin{abstract}
The current state-of-the-art of population synthesis is reviewed. The field is currently undergoing
major revisions with the recognition of several key processes as new critical ingredients. Stochastic
effects can artificially enhance or suppress certain evolutionary phases and/or stellar mass regimes 
and introduce systematic biases in, e.g., the determination of the stellar initial mass function.
Post-main-sequence evolution is often associated with irregular variations of stellar properties on 
ultra-short time-scales. Examples are asymptotic giant branch stars and luminous blue variables, both of which
are poorly treated in the models. Stars rarely form in isolation, and the fraction of truly single
stars may be very small. Therefore, stellar multiplicity must be accounted for since many systems
will develop tidal interaction over the course of their evolution. Last but not least, stellar
rotation can drastically increase stellar temperatures and luminosities, which in turn leads to
revised mass-to-light ratios in population synthesis models.

\keywords{
binaries: close,
stars: evolution,
HII regions,
galaxies: fundamental parameters,
galaxies: photometry,
galaxies: starburst,
galaxies: stellar content
}
\end{abstract}

\firstsection % if your document starts with a section,
\section{From Fundamental Data to Cosmology}

Stellar population synthesis is at the heart of spectral energy distribution (SED) studies of galaxies.
The very nature of population synthesis makes it a rather multi-disciplinary research area. It is built
on our knowledge of stellar astrophysics, which in turn relies on fundamental data derived from
laboratory physics, such as atomic line data and nuclear reaction rates. On the other hand, those
studying galaxy evolution trust population synthesis models for the interpretation of galaxy colors and
their evolution with cosmic time. Ultimately, cosmological models often rest on our faith in population 
synthesis. The chain of assumptions that must be made is usually --- subconsciously --- assigned errors
which hierarchically decrease from the fields of cosmology over stellar astrophysics to laboratory 
data. One goal of this review is to raise the awareness of radically new developments and uncertainties
in hitherto considered ``well-known" areas and to foster feedback between the stellar and cosmological 
communities.

\section{The Basics of Population Synthesis}

Population synthesis aims to reproduce the observed galaxy SED from a set of pre-specified input
parameters. Evolutionary spectral synthesis is a special, widely used case of population synthesis, whose goal is to
reproduce observed spectra self-consistently from the star-formation history of a galaxy
and from stellar evolution models. This method was pioneered by Tinsley (1980) after the advent
of modern stellar evolutionary tracks in the Hertzsprung-Russell diagram (HRD).
Evolutionary synthesis has relatively few free parameters but its success depends on the reliability of
the adopted stellar models. Therefore, assessing these assumptions becomes critical.

The astrophysical ingredients entering evolutionary synthesis are as follows: 
(i) Quantities related to the star-formation process, i.e., the star-formation
rate and its evolution with time and the stellar mass spectrum at birth, also known as
the initial mass function (IMF). See, e.g., the conference proceedings of Treyer et al. (2011). 
(ii) Once stars have formed, stellar evolution models provide a prescription for the 
variation of the stellar luminosity ($L$), effective temperature ($T_{\rm eff}$), and mass ($M$) as a function of 
chemical composition, initial mass, and time. (iii) Spectral libraries describe the spectrum of each star for
any given ($L$, $T_{\rm eff}$, $M$). (iv) Second-order effects, such as dust attenuation or geometric effects are often accounted for in a very
approximate way or even neglected altogether. 

Several widely used synthesis models are made available to the community via dedicated 
web sites. Among the most popular packages are GALEV (Schulz et al. 2002; Kotulla et al. 2009), GALEXEV
(Bruzual \& Charlot 1993; 2003), PEGASE (Fioc \& Rocca-Volmerange 1997; Le Borgne et al. 2004), and 
Starburst99 (Leitherer et al. 1999; Leitherer \& Chen 2009). In general, the predictions made
by different synthesis codes are in good agreement. When discrepancies are found, they can usually
be understood in terms different intrinsic input parameters and/or treatment of peculiar stellar
evolutionary phases, such as the asymptotic giant branch (AGB) phase (V\'azquez \& Leitherer 2005; Conroy 
\& Gunn 2010).
   
The mentioned synthesis packages restrict their output to models for the stellar luminosities, with some generic
contribution by nebular emission. More specialized applications generate panchromatic galaxy SEDs accounting
for the stellar as well as the gas emission using full photo-ionization modeling (Dopita et al. 2005; 2006a; 2006b;
Groves et al. 2008). SEDs with self-consistent inclusion of dust were first introduced by Dwek \& Scalo (1980) and refined
most recently by Dwek \& Cherchneff (2011). Chemical evolution models for the full life-cycle of the major elements over
the cosmic evolution of galaxies were discussed by Matteucci (2009; 2010).

Rather than reviewing these ``traditional'' synthesis models
I will focus on some recent, new developments which may challenge well-established results: stochasticity,
peculiar evolutionary phases, stellar multiplicity, and stellar rotation. The 
importance of these mechanisms and concepts had been recognized for quite some time but only in recent years
have they been implemented in population synthesis codes for {\em quantitative} comparison with observations.

\section{Stochasticity}

\begin{figure}[t]
% \vspace*{-2.0 cm}
\begin{center}
 \includegraphics[width=0.5\textwidth]{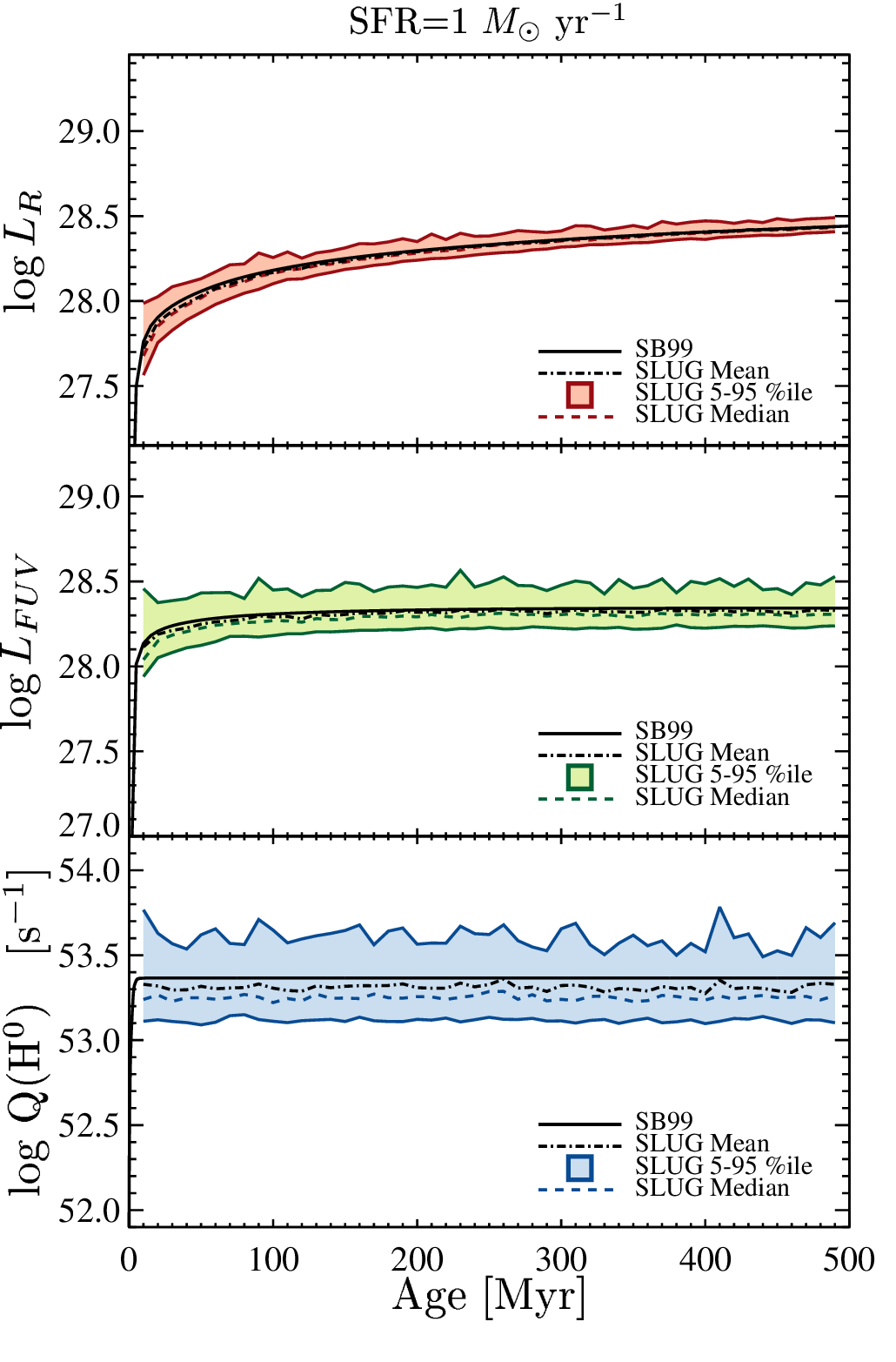}\includegraphics[width=0.5\textwidth]{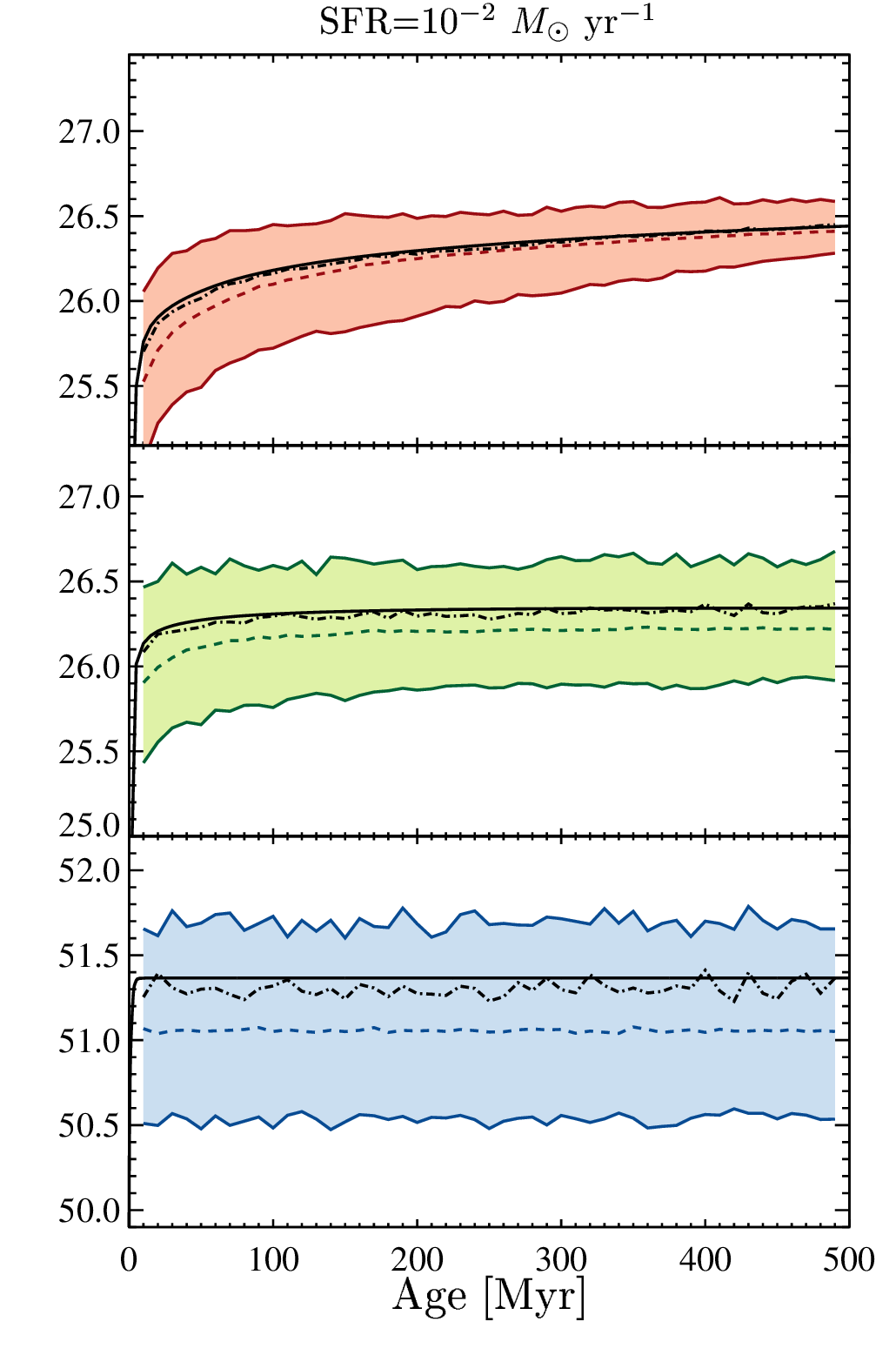}
% \vspace*{-1.0 cm}
 \caption{R-band, far-ultraviolet (UV), and ionizing luminosities vs. time for galaxies with constant star-formation rates of 1 (left) and 0.01 (right)
 $M_\odot$~yr$^{-1}$ assuming clustered star formation. Starburst99 models (solid black lines) are compared with simulations from SLUG. The SLUG models are represented
by their mean (dash-dotted line), median (dashed line) and 5 -- 95 percentile range (shaded region). From da Silva et al. (2011).}
   \label{fig1}
\end{center}
\end{figure}

Traditional population synthesis models scale all population properties by the star-formation rate. This
approach is strictly valid only in the limit of a stellar population containing an infinite
number of stars. Real star clusters or galaxies, however, contain a finite number of stars. Furthermore,
most of the light is sometimes provided by very few bright stars, in particular in the near-infrared (IR). The so-called
stochastic fluctuations in the integrated population properties are the result of the random presence of these luminous 
stars. Depending on population age and wavelength domain, stochastic effects can become important for
masses as high as $\sim$10${^5}$~$M_\odot$ (Bruzual 2002; Cervi\~no et al. 2002; Fouesneau \& Lan{\c c}on 2010).

da Silva et al. (2011) developed and released SLUG, a new code to ``Stochastically Light Up Galaxies''. SLUG
synthesizes stellar populations with a Monte Carlo technique to treat stochastic sampling properly. Included are
the effects of clustering, the stellar IMF, star-formation history, stellar
evolution, and cluster disruption. Their code has been built using the same ingredients as in Starburst99, thereby allowing
the immediate separation of stochastic effects. An example is shown in Fig.~\ref{fig1}. This figure highlights
the increased scatter and systematic offsets due to stochasticity and clustering at low star-formation rates.

GALEX UV imagery suggests that star-formation rates inferred from H$\alpha$ in galactic environments characterized by low stellar and gas densities tend to be less than those based on the UV luminosity (Lee et al. 2009). The origin of the discrepancy is still under debate. One possible explanation is that the stellar IMF is systematically deficient in high-mass stars in such environments. However, effects of uncertainties in the stellar evolutionary tracks and model atmospheres, uncertain metallicities, non-constant star-formation histories, leakage of ionizing photons, departures from Case B recombination, dust attenuation, and finally stochasticity in the formation of high-mass stars need to be considered.

Eldridge (2011) found that the scatter and variation of the H$\alpha$- and UV derived star-formation rates
are less dependent on the IMF but more on the star-formation history of each individual galaxy. The
general trend is that those systems with weaker star-formation in the last 10 Myr have relatively lower
H$\alpha$ rates, while those with most of the star formation in the
last 10 Myr display the opposite behavior even at low
mean star-formation rates. Therefore any simulation attempting to predict
the properties of a sample of galaxy must take into account
the stochastic nature of star-formation. If there are only a few clusters, the appearance of
the galaxy-wide stellar population may be very
different from what might be expected for a simple stellar population
with a smooth star-formation history.

\section{Rapid Evolutionary Phases}

\begin{figure}[t]
% \vspace*{-2.0 cm}
\begin{center}
\includegraphics[width=0.8\textwidth]{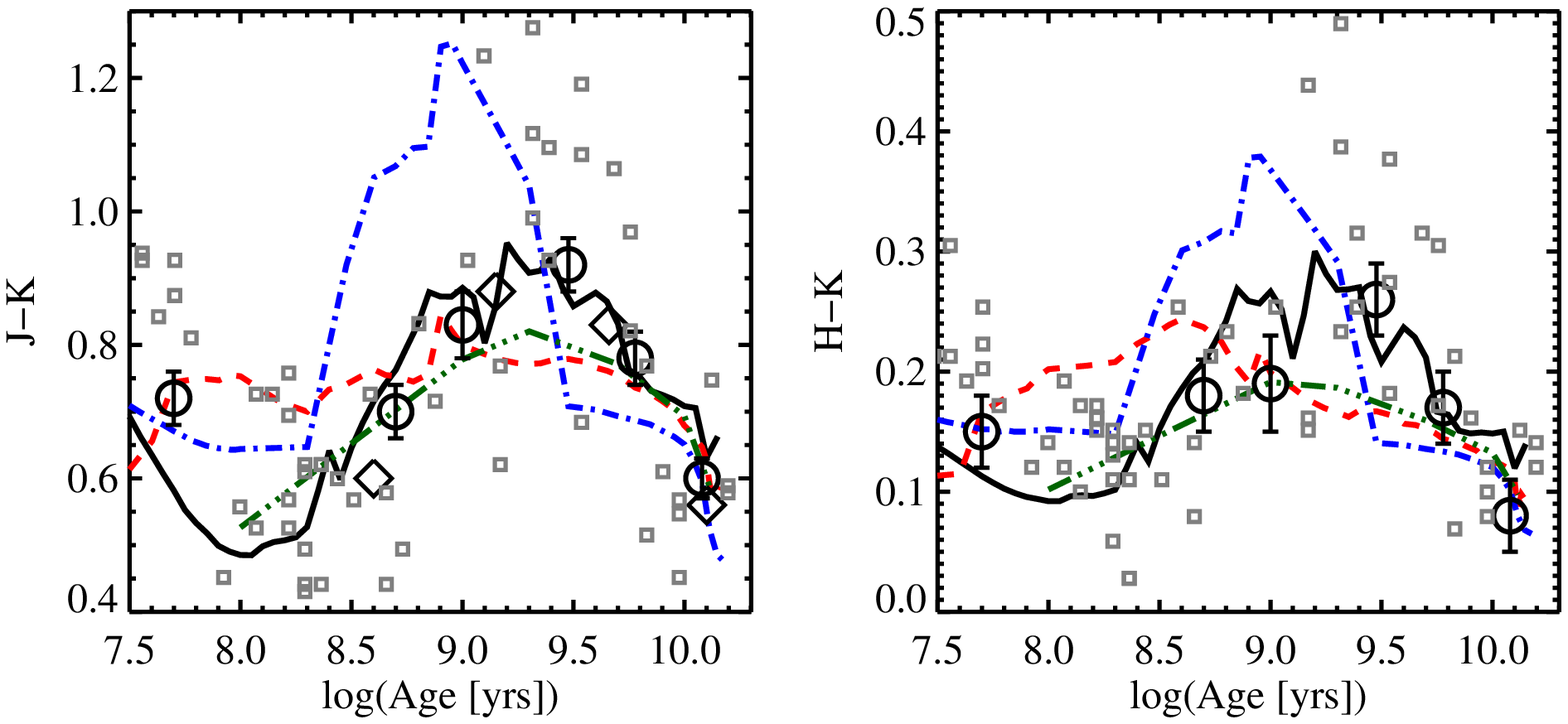} 
% \vspace*{-1.0 cm}
 \caption{Comparison of observed and modeled near-IR colors as a function of age. The data are derived from star clusters in the 
Magellanic Clouds. The models include
the predictions from Maraston (2005; dot-dashed lines), Bruzual \& Charlot (2003; triple-dot-dashed lines) and two FSPS models (dashed and solid lines).
The observed data are individual measurements (squares) and averages over many star clusters (circles). From Conroy \& Gunn (2010).}
   \label{fig2}
\end{center}
\end{figure}

Rapid evolutionary phases pose particular challenges for population synthesis. Thermally pulsing AGB stars are
especially notorious (Maraston 2011). Their uncertainties in stellar evolution modeling arise
from short pulsation time-scales ($\sim$10$^{4}$~yr), the double-shell burning, and strong mass loss in this phase. The spectral modeling
during the carbon-rich phase adds to the complexity. Yet AGB stars can often make
a significant contribution to the galaxy SED in the optical and near-IR because the
AGB phase is bright and occurs around $\sim$1~Gyr when the overall stellar population has already 
faded. Different population synthesis models disagree quite severely in their 
predicted luminosity contribution from thermally pulsing AGB stars. The Flexible
Stellar Population Synthesis (FSPS) technique of Conroy et al. (2009; 2010) and 
Conroy \& Gunn (2010) is particularly well suited for addressing this issue. FSPS
is a novel Monte-Carlo-type approach that is capable of flexibly handling various uncertain aspects of stellar
evolution such as, e.g., thermally pulsing AGB stars, the horizontal branch, or blue stragglers.
The technique allows one to marginalize over uncertain evolutionary aspects
in order to understand the full uncertainties associated with the
physical properties of galaxies including stellar masses, ages, metallicities, and star-formation rates. An example is in Fig.~\ref{fig2} where
the AGB phase as implemented in different evolutionary tracks is tested. The conclusion is that models
generally overestimate this phase, which is consistent with the result of Kriek et al. (2010) who used a sample of $\sim$60 post-starburst galaxies at $z \approx 1.6$ in the NEWFIRM medium-band survey to assess different models. The AGB contribution to the integrated SED is a factor of $\sim$3 lower than predicted. This could be due to lower bolometric luminosities, shorter lifetimes, and/or heavy dust obscuration of AGB stars.

AGB stars affect population synthesis models both directly (via their luminosity contribution) and indirectly (via their strong mass loss
influencing stellar evolution). Luminous blue variables (LBVs) mirror AGB stars in their eruptions and their high mass-loss
rates. They introduce uncertainties in population synthesis similar to those of AGB stars. They generally do not contribute 
significantly to the galaxy SED but their mass ejections are decisive for the final evolution of massive stars. It has recently been
recognized that most mass loss in massive stars prior to the Wolf-Rayet phase occurs in LBVs, whereas steady stellar winds close to
the main-sequence are less important (Smith 2010). LBVs belong to a diverse class of transients in the upper HRD spanning a wide range
of outburst luminosity which ranges from that of classical novae to that of core-collapse supernovae (Smith et al. 2011). The ubiquity of eruptive
phases and highly time-variable evolutionary phase introduce formidable challenges to population synthesis.

\section{Stellar Multiplicity}

\begin{figure}[t]
% \vspace*{-2.0 cm}
\begin{center}
 \includegraphics[width=0.75\textwidth]{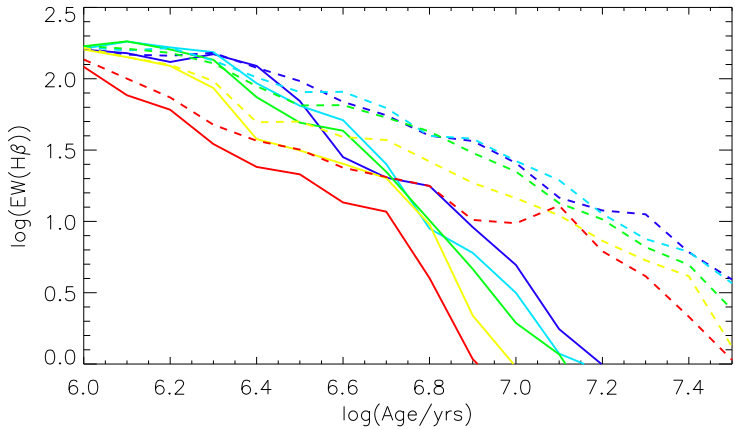} 
% \vspace*{-1.0 cm}
 \caption{H$\beta$ equivalent width vs. time for a series of instantaneous
burst models generated with the BPASS code. Solid and dashed lines indicate
populations without and with close binary stars, respectively. Both model
families with and without binaries are shown for five different metallicities,
higher metallicities generally having higher equivalent widths. From Eldridge \& Stanway (2009).}
   \label{fig3}
\end{center}
\end{figure}

Massive stars are known for their high-degree of multiplicity. Most OB stars are found in binaries and multiple systems. Combining information from these various 
surveys, Mason et al. (2009) found a minimum multiplicity fraction close to 70\%.
Even bona fide single field stars may have been part of a multiple system in the past,
then ejected by a supernova explosion or by dynamical interaction. While multiplicity deserves studies by its own
right, neglecting this fundamental property can be a serious omission in population synthesis (Vanbeveren 2010). In the
following we will not focus on the well-known vertical main-sequence displacement in the color-magnitude diagram caused 
by binaries but rather on the modified stellar evolution of close binary system and their consequences for population synthesis. 
Relevant in this context is the fraction of {\em close} binaries, i.e., those binaries whose evolution differs from that of single stars. Sana
\& Evans (2011) estimate a spectroscopic binary fraction of close to 50\%, with a significant fraction likely to be interacting.

de Mink et al. (2009) demonstrated that  the evolution of massive, close binaries still on the
main-sequence sharply differs from single-star evolution. Tidal effects can lead to significant spin-up and high rotational velocities. Mixing
processes induced by rotation may be so efficient that helium produced in the center is mixed throughout the envelope. Such
stars evolve almost chemically homogeneously. If the metallicity is low, they remain blue and compact, while they gradually evolve
into Wolf-Rayet stars and possibly into progenitors of long $\gamma$-ray bursts.

Alternatively, at low rotation velocities, the distinguishing property of binary evolution will be Roche-lobe overflow, which typically
occurs off the main-sequence. Depending on the system properties, the primary will transfer material to the secondary, which will accrete
material and gain mass. The crucial point for population synthesis is the ``rejuvenation effect'' (van Bever \& Vanbeveren 1998; Eldridge
\& Stanway 2009): the previously less massive secondary will become more massive and more luminous and will therefore appear younger. This is
illustrated in Fig.~\ref{fig3}, which shows the evolution of the nebular H$\beta$ emission-line equivalent width of single 
stellar populations. The models were obtained with the BPASS (Binary Population and Spectral Synthesis) code of Eldridge et al. (2008), which
accounts for the evolutionary effects of massive binaries. Compared with single-star models, binary populations are predicted to have
larger H$\beta$ equivalent width at older ages due to the rejuvenation effect. H$\beta$ is commonly used as an age indicator in H~II
regions (Stasi{\'n}ska et al. 2001) and would require recalibration. For instance, in single-star models, a large H$\beta$ value at older
ages would be incompatible with an instantaneous burst and suggest ongoing star formation. Binary models for single stellar populations, however, 
might still produce large enough H$\beta$ emission at older ages.

\section{Stellar Rotation}

Massive stars exhibit significant helium and nitrogen enrichment on the surface while still on the main-sequence (Hunter et al. 2009).
These telltale signs of nuclear processing require an efficient transportation mechanism to expose materials from the convective core. While
mass removal by stellar winds had previously been thought to assume this role, the main-sequence mass-loss rates are now considered too low (see
Section~4) to make this a viable option. Alternatively, mixing, mass transfer, and mass loss in close binaries might cause the enrichment.
However, it seems unlikely that the evolution of {\em all} massive stars is driven by binarity. Many stars with surface enrichment are also
fast rotators. While the relation between nitrogen overabundance and rotation velocity is complex and far from unique, there is
consensus that rotation is an important, if not the prime driver of mixing in massive stars (Brott et al. 2011a).
Stellar evolution with rotation has been modeled over the past decade, but it is only now that {\em quantitative} evolutionary tracks with
rotation have become available for implementation of population synthesis (Brott et al. 2011a; Ekstr\"om et al. 2011).

\begin{figure}[t]
% \vspace*{-2.0 cm}
\begin{center}
 \includegraphics[angle=-90,width=0.85\textwidth]{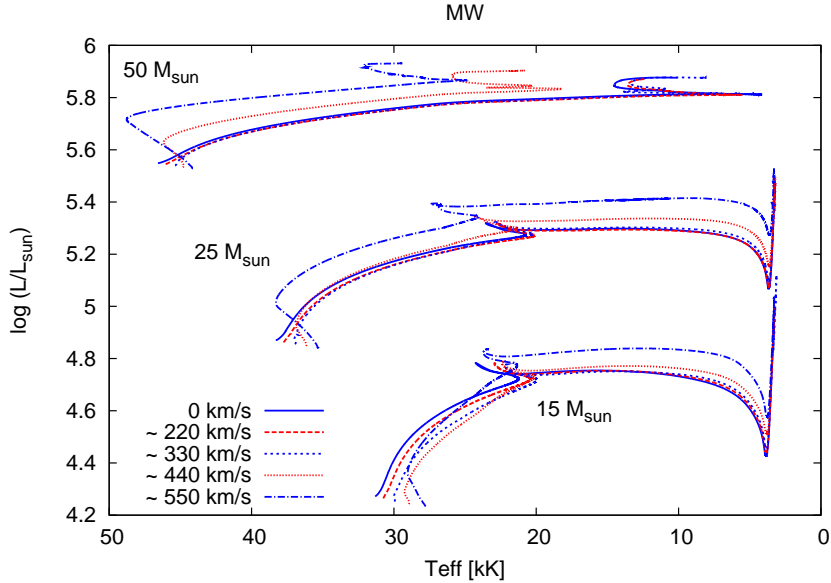} 
% \vspace*{-1.0 cm}
 \caption{Evolutionary tracks for rotating stars with initial masses of 15, 25, and 50~$M_\odot$ having solar chemical composition. Line types indicate different 
initial rotation velocities on the zero-age main-sequence. From Brott et al. (2011b).}
   \label{fig4}
\end{center}
\end{figure}

Fig.~\ref{fig4} illustrates the principal effects caused by rotation. At high masses ($M > 20 M_\odot$), the dominant mechanism is the increase of the 
convective core and the reduced surface opacity due to the He enhancement, resulting in a higher $L$ and higher $T_{\rm eff}$, respectively. (Recall that H is
the major opacity source and any decrease of its relative abundance by mixing lowers the opacity and therefore increases $T_{\rm eff}$.) Therefore
tracks with rotation tend to fall above and to the left of non-rotating tracks. At lower masses (e.g., the track for $M = 15~M_\odot$ in Fig.~\ref{fig4}), 
these effects become less important and are surpassed by the impact of the centrifugally induced radius increase and the associated lower $T_{\rm eff}$.
At masses below $\sim$2~$M_\odot$ no rotational effects are expected since low-mass stars arrive on the zero-age main-sequence with essentially no
velocity because of magnetic braking.

We implemented the new suite of 48 different stellar evolutionary tracks, both rotating and non-rotating, by Ekstr\"om et al. (2011) in Starburst99. The grid has
solar chemical composition and covers the full mass range from 0.8 to 120~$M_\odot$. The rotating models start on the zero-age
main-sequence with a rotation velocity of 40\% of the equatorial break-up velocity. Guidance on the choice of the initial rotation velocity is 
provided by measurements in B stars but data for O stars are still scarce. Clearly this parameter must be considered with care. 
The evolution is computed until the end of the central carbon-burning phase, the early AGB phase, or the core helium-flash for the massive, intermediate, and low-mass stars, respectively. 
  V\'azquez et al. (2007) used an earlier, preliminary version of these models to test the impact of stellar rotation on synthetic
population properties. The new model suite will allow us to verify and extend the previous work (Levesque et al. 2012, in preparation).

\begin{figure}[t]
% \vspace*{-2.0 cm}
\begin{center}
\includegraphics[width=0.5\textwidth]{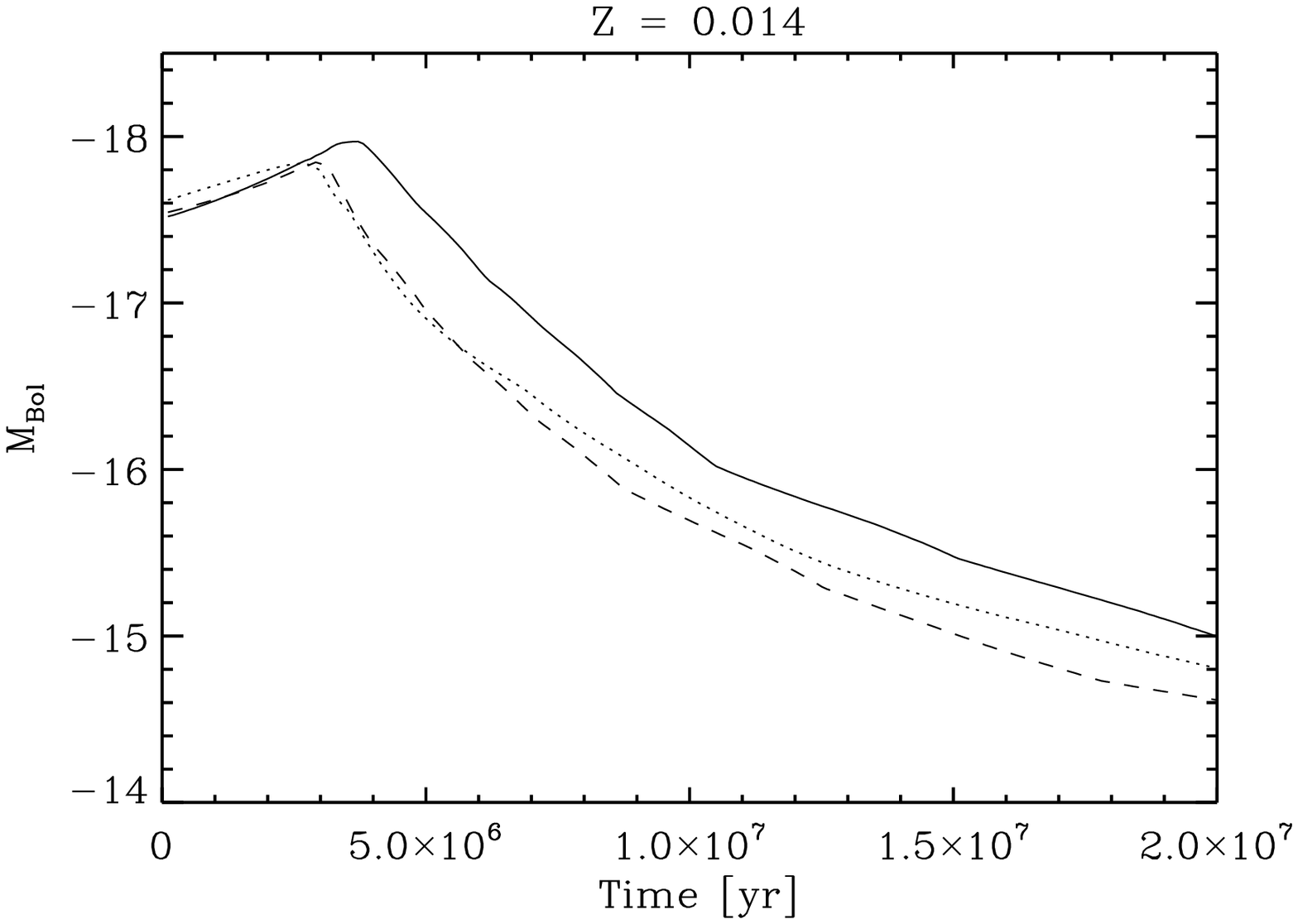}\includegraphics[width=0.5\textwidth]{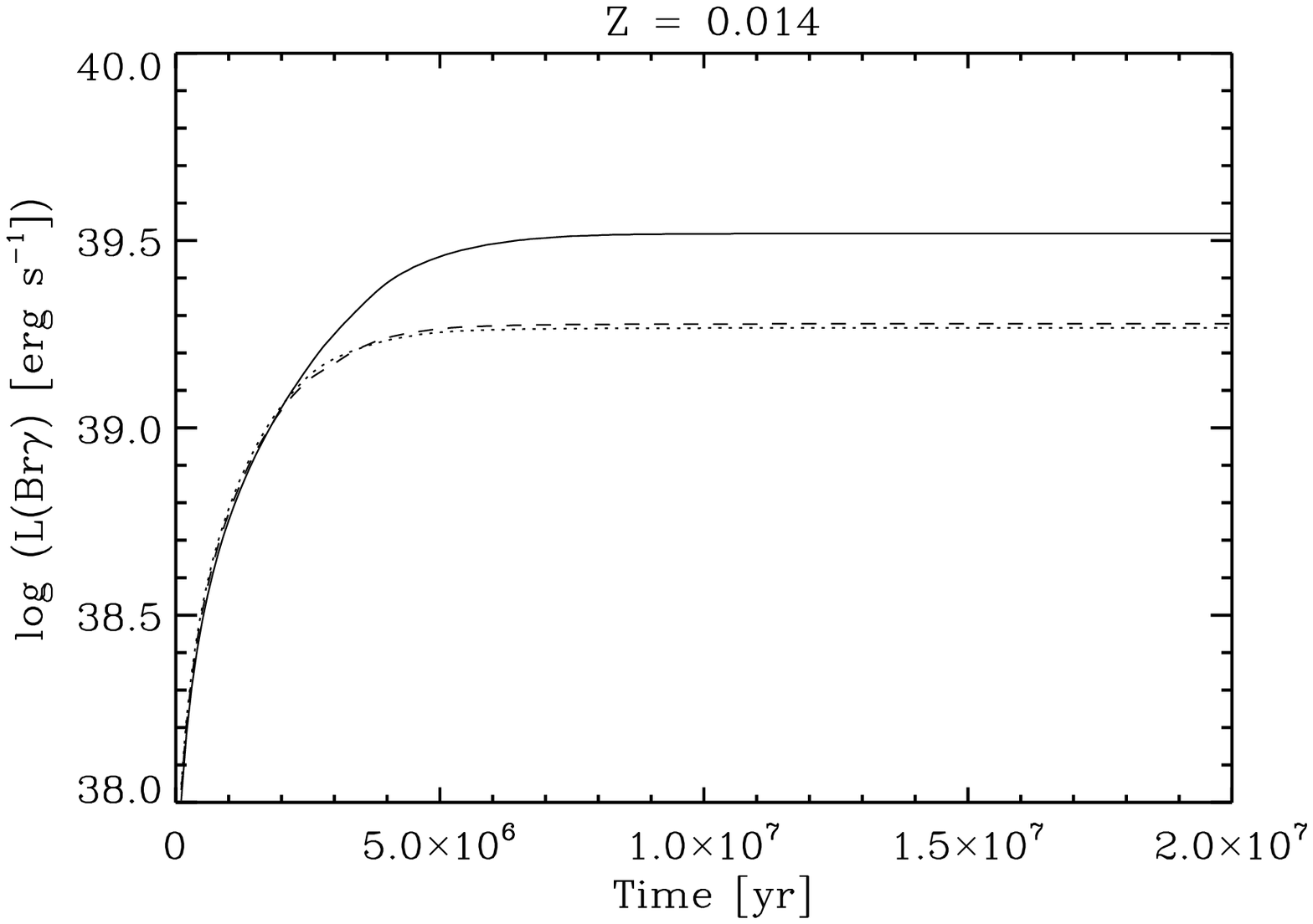}

% \vspace*{-1.0 cm}
 \caption{Comparison of stellar population properties resulting from evolutionary tracks with and without rotation. Left: bolometric
luminosity vs. time for a single stellar population with a mass of $10^6$ $M_\odot$. Right: Br$\gamma$ luminosity for continuous star formation at a rate
of 100~$M_\odot$~yr$^{-1}$. Solid lines: new tracks with 
rotation; dashed: new tracks with zero rotation velocity; dotted: previous generation of Geneva tracks released 1992 -- 1994. The models are for
solar chemical composition and use a standard Kroupa IMF between 0.1 and 100~$M_\odot$.}
   \label{fig5}
\end{center}
\end{figure}

Rotation increases both $L$ and $T_{\rm eff}$ for stars more massive than $\sim$20~$M_\odot$. This effect sets in very early on the main-sequence. As a 
result, a stellar population containing hot, massive stars is predicted to be more luminous for a given mass, and its SED is shifted to shorter wavelengths.
Therefore the most dramatic changes with respect to prior models occur at the short-wavelength end of the SED. Examples are reproduced in Fig.~\ref{fig5},
which compares $M_{\rm Bol}$ and the luminosity of the nebular Br$\gamma)$ resulting from models with and without rotation. The single stellar population models with rotation for $M_{\rm Bol}$ become
more luminous by $\sim$0.4~mag (left part of Fig.~\ref{fig5}). The right part of Fig.~\ref{fig5} shows the Br$\gamma$ luminosity for
an equilibrium population. Br$\gamma$ is often used to determine star-formation rates in obscured galaxies, such as ultra-luminous IR galaxies. For
this specific example, the rates following from the models with and without rotation would be 100 and 175~$M_\odot$~yr$^{-1}$, respectively.  
The extreme UV increases by a factor of a few in the hydrogen ionizing continuum and by several orders of magnitude in 
the neutral and ionized helium continua with the new models. The predictions for the latter need careful testing, as the photon
escape fraction crucially depends on the interplay between the stellar parameters supplied by the evolution
models and the radiation-hydrodynamics of the atmospheres. In contrast, the escape of the hydrogen ionizing
photons has little dependence on the particulars of the atmospheres and consequently is a relatively safe 
prediction.

\section{The New Standard}

Stochasticity, exotic stellar evolutionary phases, multiplicity, and rotation are now considered essential ingredients in population
synthesis models. A new suite of population synthesis models which account for these effects and provide
quantitative predictions have become available in the very recent past. Somewhat frustratingly, implementation of these processes cannot rely on first principles, which essentially adds new
free parameters requiring empirical calibration. Care is therefore required when the first model generation is used to analyze galaxy SEDs. 
The community is encouraged to perform extensive testing and help making these new synthesis models the new standard.

\acknowledgments

Claus Leitherer acknowledges travel support from the Director's Discretionary Research Fund at STScI.


\begin{thebibliography}{}

\bibitem[Brott et 
al.(2011)]{2011A&A...530A.115B} Brott, I., de Mink, S.~E., Cantiello, M., et al.\ 2011b, {\it A\&A}, 530, A115 


\bibitem[Brott et 
al.(2011)]{2011A&A...530A.116B} Brott, I., Evans, C.~J., Hunter, I., et al.\ 2011a, {\it A\&A}, 530, A116 


\bibitem[Bruzual A.(2002)]{2002IAUS..207..616B} Bruzual A., G.\ 2002, in IAU Symp. 207, Extragalactic Star Clusters,
  eds.  D. Geisler, E. K. Grebel, \& D. Minniti (San Francisco: ASP), 616 

\bibitem[Bruzual 
A.~\& Charlot(1993)]{1993ApJ...405..538B} Bruzual A., G., \& Charlot, S.\ 1993, {\it ApJ}, 405, 538 

\bibitem[Bruzual 
\& Charlot(2003)]{2003MNRAS.344.1000B} ------.\ 2003, {\it MNRAS}, 344, 1000 

\bibitem[Cervi{\~n}o et 
al.(2002)]{2002A&A...381...51C} Cervi{\~n}o, M., Valls-Gabaud, D., Luridiana, V., \& Mas-Hesse, J.~M.\ 2002, {\it A\&A}, 381, 51 

\bibitem[Conroy 
\& Gunn(2010)]{2010ApJ...712..833C} Conroy, C., \& Gunn, J.~E.\ 2010, {\it ApJ}, 712, 833 

\bibitem[Conroy et al.(2009)]{2009ApJ...699..486C} Conroy, C., Gunn, J.~E., 
\& White, M.\ 2009, {\it ApJ}, 699, 486 

\bibitem[Conroy et al.(2010)]{2010ApJ...708...58C} Conroy, C., White, M., 
\& Gunn, J.~E.\ 2010, {\it ApJ}, 708, 58 

\bibitem[da Silva et al.(2011)]{2011arXiv1106.3072D} da Silva, R.~L., 
Fumagalli, M., \& Krumholz, M.\ 2011, arXiv:1106.3072 

\bibitem[de Mink et 
al.(2009)]{2009A&A...497..243D} de Mink, S.~E., Cantiello, M., Langer, N., et al.\ 2009, {\it A\&A}, 497, 243 

\bibitem[Dopita et al.(2006b)]{2006ApJS..167..177D} Dopita, M.~A., Fischera, 
J., Sutherland, R.~S., et al.\ 2006b, {\it ApJS}, 167, 177 


\bibitem[Dopita et al.(2006b)]{2006ApJ...647..244D} Dopita, M.~A., Fischera, 
J., Sutherland, R.~S., et al.\ 2006a, {\it ApJ}, 647, 244 


\bibitem[Dopita et al.(2005)]{2005ApJ...619..755D} Dopita, M.~A., Groves, 
B.~A., Fischera, J., et al.\ 2005, {\it ApJ}, 619, 755 

\bibitem[Dwek 
\& Cherchneff(2011)]{2011ApJ...727...63D} Dwek, E., \& Cherchneff, I.\ 2011, {\it ApJ}, 727, 63 

\bibitem[Dwek 
\& Scalo(1980)]{1980ApJ...239..193D} Dwek, E., \& Scalo, J.~M.\ 1980, {\it ApJ}, 239, 193 

\bibitem[Ekstr{\"o}m et al.(2011)]{2011arXiv1110.5049E} Ekstr{\"o}m, S., 
Georgy, C., Eggenberger, P., et al.\ 2011, arXiv:1110.5049 

\bibitem[Eldridge(2011)]{2011arXiv1106.4311E} Eldridge, J.~J.\ 2011, 
arXiv:1106.4311 

\bibitem[Eldridge et al.(2008)]{2008MNRAS.384.1109E} Eldridge, J.~J., 
Izzard, R.~G., \& Tout, C.~A.\ 2008, {\it MNRAS}, 384, 1109 

\bibitem[Eldridge 
\& Stanway(2009)]{2009MNRAS.400.1019E} Eldridge, J.~J., \& Stanway, E.~R.\ 2009, {\it MNRAS}, 400, 1019 


\bibitem[Fioc 
\& Rocca-Volmerange(1997)]{1997A&A...326..950F} Fioc, M., \& Rocca-Volmerange, B.\ 1997, {\it A\&A}, 326, 950 

\bibitem[Fouesneau 
\& Lan{\c c}on(2010)]{2010A&A...521A..22F} Fouesneau, M., \& Lan{\c c}on, A.\ 2010, {\it A\&A}, 521, A22 

\bibitem[Groves et al.(2008)]{2008ApJS..176..438G} Groves, B., Dopita, 
M.~A., Sutherland, R.~S., et al.\ 2008, {\it ApJS}, 176, 438 

\bibitem[Hunter et 
al.(2009)]{2009A&A...496..841H} Hunter, I., Brott, I., Langer, N., et al.\ 2009, {\it A\&A}, 496, 841 


\bibitem[Kotulla et al.(2009)]{2009MNRAS.396..462K} Kotulla, R., Fritze, 
U., Weilbacher, P., \& Anders, P.\ 2009, {\it MNRAS}, 396, 462 

\bibitem[Kriek et al.(2010)]{2010ApJ...722L..64K} Kriek, M., Labb{\'e}, I., 
Conroy, C., et al.\ 2010, {\it ApJ}, 722, L64 


\bibitem[Le Borgne et 
al.(2004)]{2004A&A...425..881L} Le Borgne, D., Rocca-Volmerange, B., Prugniel, P., et al.\ 2004, {\it A\&A}, 425, 881 

\bibitem[Lee et al.(2009)]{2009ApJ...706..599L} Lee, J.~C., Gil de Paz, A., 
Tremonti, C., et al.\ 2009, {\it ApJ}, 706, 599 

\bibitem[Leitherer 
\& Chen(2009)]{2009NewA...14..356L} Leitherer, C., \& Chen, J.\ 2009, {\it New Astronomy}, 14, 356 

\bibitem[Leitherer et al.(1999)]{1999ApJS..123....3L} Leitherer, C., 
Schaerer, D., Goldader, J.~D., et al.\ 1999, {\it ApJS}, 123, 3 

\bibitem[Maraston(2005)]{2005MNRAS.362..799M} Maraston, C.\ 2005, {\it MNRAS}, 
362, 799 


\bibitem[Maraston(2011)]{2011arXiv1104.0022M} ------.\ 2011, in Why AGB Stars care about Galaxies II, 
  eds.  F. Kerschbaum, T. Lebzelter, \& B. Wing (San Francisco: ASP), in press (arXiv:1104.0022) 

\bibitem[Mason et al.(2009)]{2009AJ....137.3358M} Mason, B.~D., Hartkopf, 
W.~I., Gies, D.~R., Henry, T.~J., \& Helsel, J.~W.\ 2009, {\it AJ}, 137, 3358 


\bibitem[Matteucci(2009)]{2009AIPC.1111..143M} Matteucci, F.\ 2009, in Probing Stellar Populations out to the Distant Universe, ed. L. A. Antonelli et al. (Melville: AIP), 143 

\bibitem[Matteucci(2010)]{2010gama.conf..261M} ------.\ 2010, in
Galaxies and their Masks, eds. L. David, K. C. Freeman, \& I. Puerari (New York: Springer),  261 

\bibitem[Sana 
\& Evans(2011)]{2011IAUS..272..474S} Sana, H., \& Evans, C.~J.\ 2011, in IAU Symp. 272, Active OB Stars, eds. C.~Neiner, G.~Wade, G.~Meynet, \& G.~Peters (Cambridge: CUP),
474 

\bibitem[Schulz et 
al.(2002)]{2002A&A...392....1S} Schulz, J., Fritze-v.~Alvensleben, U., M{\"o}ller, C.~S., \& Fricke, K.~J.\ 2002, {\it A\&A}, 392, 1 

\bibitem[Smith(2010)]{2010ASPC..425...63S} Smith, N.\ 2010, in Hot and Cool: 
Bridging Gaps in Massive Star Evolution, eds. C. Leitherer, P. Bennett, P. Morris, \& J. van Loon (San Francisco: ASP), 63 


\bibitem[Smith et al.(2011)]{2011MNRAS.415..773S} Smith, N., Li, W., 
Silverman, J.~M., Ganeshalingam, M., 
\& Filippenko, A.~V.\ 2011, {\it MNRAS}, 415, 773 

\bibitem[Stasi{\'n}ska et 
al.(2001)]{2001A&A...370....1S} Stasi{\'n}ska, G., Schaerer, D., \& Leitherer, C.\ 2001, {\it A\&A}, 370, 1 

\bibitem[Tinsley(1980)]{1980FCPh....5..287T} Tinsley, B.~M.\ 1980, {\it Fund. Cosm. Phys.}, 5, 
287 

\bibitem[Treyer et al.(2011)]{2011ASPC..440.....T} Treyer, M., Wyder, T., 
Neill, J., Seibert, M., 
\& Lee, J.\ 2011, UP2010: Have Observations Revealed a Variable Upper End of the Initial Mass Function? (San Francisco: ASP)  

\bibitem[van Bever 
\& Vanbeveren(1998)]{1998A&A...334...21V} van Bever, J., \& Vanbeveren, D.\ 1998, {\it A\&A}, 334, 21 

\bibitem[Vanbeveren(2010)]{2010IAUS..266..293V} Vanbeveren, D.\ 2010, in IAU 
Symp. 266, Star Clusters: Basic Galactic Building Blocks throughout Time and Space, eds R.~de Grijs \& J.~R.~D.~L{\'e}pine (Cambridge: CUP), 293 


\bibitem[V{\'a}zquez 
\& Leitherer(2005)]{2005ApJ...621..695V} V{\'a}zquez, G.~A., \& Leitherer, C.\ 2005, {\it ApJ}, 621, 695 

\bibitem[V{\'a}zquez et al.(2007)]{2007ApJ...663..995V} V{\'a}zquez, G.~A., 
Leitherer, C., Schaerer, D., Meynet, G., 
\& Maeder, A.\ 2007, {\it ApJ}, 663, 995 






\end{thebibliography}
\end{document}